# ARTEMIS: A Mixed Analog-Stochastic In-DRAM Accelerator for Transformer Neural Networks

Salma Afifi, *Student Member, IEEE*, Ishan Thakkar, *Member, IEEE* and Sudeep Pasricha, *Fellow, IEEE*

*Abstract*—Transformers have emerged as a powerful tool for natural language processing (NLP) and computer vision. Through the attention mechanism, these models have exhibited remarkable performance gains when compared to conventional approaches like recurrent neural networks (RNNs) and convolutional neural networks (CNNs). Nevertheless, transformers typically demand substantial execution time due to their extensive computations and large memory footprint. Processing in-memory (PIM) and near-memory computing (NMC) are promising solutions to accelerating transformers as they offer high compute parallelism and memory bandwidth. However, designing PIM/NMC architectures to support the complex operations and massive amounts of data that need to be moved between layers in transformer neural networks remains a challenge. We propose ARTEMIS, a mixed analog-stochastic in-DRAM accelerator for transformer models. Through employing minimal changes to the conventional DRAM arrays, ARTEMIS efficiently alleviates the costs associated with transformer model execution by supporting stochastic computing for multiplications and temporal analog accumulations using a novel in-DRAM metal-on-metal capacitor. Our analysis indicates that ARTEMIS exhibits at least 3.0 × speedup, 1.8 × lower energy, and 1.9 × better energy efficiency compared to GPU, TPU, CPU, and state-of-the-art PIM transformer hardware accelerators.

## I. INTRODUCTION

IN recent years, the capabilities of transformer neural networks have revolutionized the landscape of artificial intelligence, eclipsing traditional architectures like recurrent neural networks (RNNs) and convolutional neural networks (CNNs) across a spectrum of sequence and vision-based tasks [1]. Renowned models such as BERT [2], ALBERT [3], and GPT-4 [4] have emerged as leading solutions in natural language processing (NLP), with unparalleled accuracies in tasks ranging from machine translation to named entity recognition and question-answering. Transformers have also demonstrated success across various visual tasks, facilitated by the implementation of vision transformers (ViT) [5].

However, as the pursuit of higher accuracies leads to the development of increasingly complex transformer neural networks, a surge in model size and parameter count has been observed. Large transformer networks, designed to capture intricate relationships within ever-expanding sequences, demand billions of parameters [3], [4]. Yet, this exponential growth in parameters is not without consequences. With each increase in model size and sequence length, the communication overhead required to move parameters between memory and compute units becomes a bigger bottleneck. Notably, the energy consumption linked to data transfers between processors and off-chip memory now exceeds that of a floating-point operation by a factor of two orders of magnitude [6].

Current ASIC and FPGA-based accelerators tailored for transformers such as [7] and [8] encounter challenges stemming from restricted parallelism and constrained off-chip memory bandwidth, thereby limiting their acceleration capabilities. In contrast, memory-based acceleration methods, such as processing in-memory (PIM) and near-memory computing (NMC), have shown great potential for speeding up transformer execution by exploiting extensive parallelism, reducing data movement costs, and offering scalable memory bandwidth [9]-[11]. In-DRAM processing, in particular, is of significant interest as it leverages and extends a ubiquitous memory component (i.e., DRAM) found in all computing platforms. However, this approach presents two major challenges: executing the intensive operations required by transformers and efficiently managing intra-memory data movement.

Transformer models involve a combination of multiply-and-accumulate (MAC) operations along with complex functions such as reduction and softmax. Previous research has integrated MAC operations within DRAM bit-cell arrays using sense amplifiers (S/A) [6]. This approach necessitates the digital implementation of MAC operations in DRAM-based PIM accelerators, which is achieved by decomposing a single MAC operation into multiple functionally complete memory operation cycles (MOCs) [6], [9]. Consequently, this approach leads to a heightened number of MOCs for MAC operations in state-of-the-art in-DRAM processing architectures, presenting a significant challenge. Moreover, implementing functions like reduction and softmax digitally within DRAM bit-cell arrays is not straightforward. Integrating embedded logic within the DRAM blocks to leverage NMC offers a possible solution to this challenge. However, this can lead to a large added area overhead. Also, effectively orchestrating dataflow, scheduling, and managing the data movement and various operations in both PIM and NMC contexts presents a complex and non-trivial task. Although the hierarchical structure of DRAM allows for highly parallelized execution across multiple DRAM banks, the movement of data is severely limited by the single bus shared among all banks. Traditional PIM methodologies typically employ layer-based dataflow schemes. However, due to the large number of parameters in transformer models, such dataflows can result in over 60% of the transformer's inference execution time consumed in data movement alone [9].

In this paper, we present ARTEMIS, the first in-DRAM accelerator that leverages mixed analog-stochastic computations for accelerating transformer neural networks. Due to the distinctive architecture of transformers and their intensive operations which involve a substantial number of

MAC computations, ARTEMIS employs stochastic computing for multiplication operations. This allows our accelerator to perform a single multiply operation in 34ns instead of 1600ns with traditional in-DRAM PIM solutions [6]. Accumulations are performed using a temporal analog accumulation approach which significantly reduces data movement overheads and enables fast and accurate successive data accumulations. To further address the intra-memory data movement bottleneck, an optimized token-based dataflow tailored for the stochastic-analog computational flow, is implemented in the software layer. With a token-based dataflow, memory resources are assigned for computations across different layers based on the input tokens [9], [10]. Accordingly, each memory bank processes and stores the intermediate results related to a specific set of tokens, thereby significantly reducing the amount of data transferred between layers. In summary, our work makes the following novel contributions:

- We design a novel in-DRAM hardware accelerator called ARTEMIS by combining principles of stochastic and analog computing, to accelerate multiple existing variants of transformer neural networks.
- We develop a novel in-DRAM analog accumulation unit by repurposing a custom metal-oxide-metal capacitor (MOMCAP) specifically for analog computing.
- We efficiently combine dataflow and control mechanisms and implement intra- and inter-bank microarchitectures to reduce data movement latencies and energy overheads.
- We demonstrate that our proposed architecture outperforms GPU, TPU, CPU, and several state-of-the-art PIM transformer neural network accelerators through a comprehensive comparison.

The rest of the paper is organized as follows. Section II provides a background on transformers, stochastic computing (SC), DRAM structures, and accelerating transformers using PIM. Section III describes the ARTEMIS framework and our optimization efforts at the device, circuit, and architecture layers. Details of the experiments conducted, simulation setup, and results are presented in Section IV. Lastly, Section V presents concluding remarks.

## II. BACKGROUND

### A. Transformer Neural Networks

The original Transformer neural network model [1] is based on $L$ layers of encoder and decoder blocks as shown in Fig. 1. The encoder transforms the input sequence into a coherent continuous representation of tokens, which is subsequently processed by the decoder. As the decoder executes, it iteratively generates a single output while incorporating the preceding outputs. The two main sub-blocks in the encoder and decoder blocks are the multi-head attention (MHA) and feed forward (FF) layers. The MHA layer implements the self-attention mechanism which has gained significant traction in sequence learning and natural language processing (NLP), particularly in scenarios where long-term memory is essential. The input to the MHA layer ($I \in R^{N \times D}$) with $N$ number of tokens, is first processed by three linear layers. The linear layers generate the query ($Q \in R^{N \times D}$), key ($K \in R^{N \times D}$), and value ($V \in R^{N \times D}$) matrices by multiplying the input matrix $I$ by weight matrices ($W^Q \in R^{D \times D}$), ($W^K \in R^{D \times D}$), and ($W^V \in R^{D \times D}$) respectively. The MHA is composed of $H$ number of heads where the dimension $D$ is split across all heads. The scaled dot-product attention is then computed as follows:

$$Head(I) = attention(Q,K,V) = softmax(QK^T/\sqrt{D}).V \quad (1)$$

The output of the MHA is the concatenation of the self-attention heads' outputs, followed by a linear layer. The FF layer consists of two dense layers with a RELU activation in between.

Newer transformer-based pre-trained language models, such as BERT and its variants [2], [3], adopt a configuration consisting solely of the transformer encoder block and a classification output layer. This block is comprised of a cascaded set of $L$ layers, followed by an FF layer, GELU activation function, and normalization layers. Similarly, the vision transformer (ViT) model also employs $L$ encoder layers, followed by a multi-layer perceptron. The ViT model inputs are sequence vectors representing an image [5].

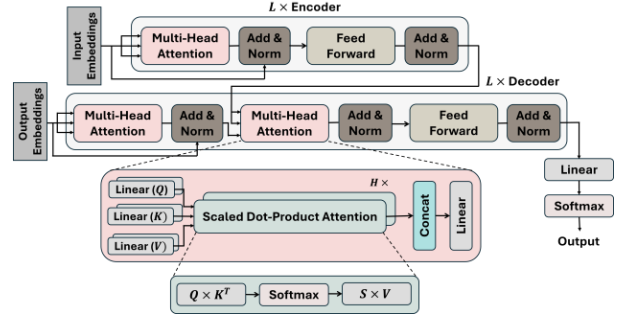

**Fig. 1.** Transformer neural network architecture overview

### B. Stochastic Computing (SC)

SC simplifies computational complexity by utilizing extended sequences of individual bits to represent numerical values. By trading off precision and representation density, SC can achieve simpler logic design and lower power consumption. Consequently, it has received a lot of attention recently in fields such as image/signal processing, control systems, deep neural networks (DNNs), and general-purpose computing [13], [14]. A system utilizing SC typically encapsulates three main steps:

**1) Data generation and representation:** SC employs extended independent bit-streams to represent real numbers probabilistically, with the occurrence rates of 1s and 0s within the streams representing the corresponding real values. Eq. (2) and (3) outline examples for stochastically representing two binary numbers.

$$X_1 = \frac{6}{10} \rightarrow x_1(stoch.) = 0110101101 \quad (2)$$
$$X_2 = \frac{4}{10} \rightarrow x_2(stoch.) = 1010010001 \quad (3)$$

Pseudo-random number generators like linear-feedback shift registers (LFSRs) are frequently employed to generate the stochastic numbers, but such methods are susceptible to random variations, leading to inaccurate computations [15]. Alternatively, stochastic representations can be obtained deterministically using a decoder or a look-up table (LUT) which eliminates the inaccuracies caused by random fluctuations or correlations between bit-streams [15].

**2) Stochastic arithmetic functions:** Stochastic computing performs computations by statistically manipulating input bit-streams. Most functions found in binary computing are also accommodated within SC [16]. However, binary computing functions that usually entail complex digital circuits can be performed with SC using simple logic gates. For example, a multiplication operation can be computed by a single AND gate using the stochastic bitstreams. Multiplying the two numbers from Eq. (2) and (3) would be computed as follows:

$$X_1 \times X_2 = x_1 \& x_2 = 0010000001 \ (= 0.2) \qquad (4)$$

The product of $X_1$ and $X_2$ is expected to yield a real value of 0.24, yet the bitwise AND operation of $x_1$ and $x_2$ produces a result of 0.2. Thus, SC can experience a degree of precision loss. Within our ARTEMIS accelerator, we introduce methodologies aimed at overcoming such inaccuracies.

**3) Stochastic to binary number conversion:** Stochastic numbers involve a storage overhead of $O(2^n)$ due to the necessity of representing an *n*-bit real value with $2^n$ bits. To mitigate this overhead, operand storage in SC typically adopts the binary format, necessitating stochastic-to-binary (S_to_B) conversions of operands. Such conversions are often performed using a popcount (PC) unit, which tallies the number of 1's in a stochastic bitstream to derive the corresponding binary value. However, PC units present several challenges due to their high area, latency, and energy overheads [17], [18]. ARTEMIS employs a low-overhead technique for S_to_B conversions.

While some prior works have started to explore SC for conventional DNN acceleration [13], [19], to the best of our knowledge, ARTEMIS represents the first architecture that tailors SC for accelerating transformer neural network models.

*C. DRAM Structure and Operation*

A DRAM chip features a hierarchical architecture consisting of banks, subarrays, and tiles. Within each subarray, there exists a two-dimensional 2D array of DRAM cells, each comprising an access transistor and a capacitor (1T1C). These subarrays are further divided into smaller tiles. The local bit-line, which encompasses multiple cells, is linked to an S/A that actively manipulates the charge while serving as a row buffer [20]. The baseline memory framework utilized in this work is Samsung's high-bandwidth memory (HBM) [12]. HBM usually comprises several stacks where each stack consists of a 4-layer HBM chip. These stacks consist of multiple DRAM slices positioned atop the base die, enabling significantly enhanced bandwidth and reduced access latency compared to traditional 2D DRAM configurations. Each chip is further divided into channels and each channel is composed of several DRAM banks.

A read operation in DRAM involves three phases: *pre-charge*, *activate*, and *restore*. During pre-charge, bit-lines are set to $\frac{Vdd}{2}$. In the subsequent activate phase, bit-lines are released while the target cells are accessed. Charge is then distributed between the cell and bit-line parasitic capacitance. The S/A engages to detect and amplify the subtle voltage variation. The amplified voltage variation is then restored to the target cells in the restore phase. In a write operation, S/As read and amplify data from the DRAM chip's internal bus, which is written to the target cells during the restore phase.

*D. Memory-based Computing*

Memory-based computing systems have received significant attention from both industry and academia. Such systems can be broadly categorized into PIM and NMC architectures. PIM embeds logic directly within the memory arrays, allowing it to perform computations on the stored data without notable data movement. This is enabled through utilizing the inherent operations already performed within the memory arrays (i.e., read and write) [21]. Meanwhile, NMC integrates logic in proximity of the memory system [22]. This can entail placing compute units in the HBM's logic die [23], in near-bank I/O, or in the near-subarray circuits inside the memory bank [24]. NMC typically incurs a higher area overhead, but it still reduces the necessity for data movement by performing computations closer to the data storage location, without altering the tile structure. Despite presenting some manufacturing challenges, recent advancements in DRAM die-stacking technology, such as HBM, have mitigated various concerns about practicality and cost. Moreover, as PIM logical operations exploits the intrinsic operation of the DRAM arrays, it can be integrated with minimal hardware modifications and manufacturing cost [25].

While DRAM-based in-memory computing has been widely explored, alternative memory technologies have also received much attention. For example, recent studies have shown that some emerging nonvolatile memory technologies, including ReRAM and phase change memory possess capabilities extending beyond mere storage functions. These technologies can perform logic operations, supporting both computation and memory tasks, thereby facilitating PIM architecture development [26]. Accordingly, several previous works have proposed utilizing such technologies for accelerating DNNs, including CNNs [27], RNNs [26], and transformers [11]. However, such architectures introduce a distinct set of challenges, e.g., ReRAM cells suffer from reliability issues [27]. ARTEMIS therefore leverages the prevalent and ubiquitous DRAM technology for computational tasks while integrating PIM and NMC principles. This enables rapid and energy-efficient acceleration of transformer neural networks.

In-DRAM PIM computing approaches integrate processing units within DRAM subarrays, leveraging the inherent mechanism of a DRAM read operation, discussed earlier. Through the utilization of RowClone [39], data transfer between different DRAM rows is achieved by concurrently activating the target row while restoring data to the original row. This process involves two consecutive activations followed by the pre-charge stage, known as the activate-activate-precharge (AAP) primitive [20]. Each AAP cycle corresponds to one memory operation cycle (MOC). Subsequent studies expanded upon this approach to incorporate fundamental functions within DRAM subarrays. For instance, Ambit [20] concurrently activates three DRAM rows to execute bulk bitwise AND and OR operations in 3 MOCs, while ROC [30] employs only two DRAM rows with an additional diode placed between each two bit-cells situated. This allows ROC to perform AND and OR operations in only 2 MOCs.

*E. PIM for Transformer Neural Network Acceleration*

Memory-based PIM hardware accelerator designs have been extensively explored for traditional DNNs such as CNNs [6], [13], [19]. Nevertheless, extending such architectures to transformer models can be inefficient. This is due to two main aspects inherent to transformer models: the unique and intensive computations within the transformer layers, and the massive amount of data that needs to be moved between those layers. Conventional PIM systems implement arithmetic functions digitally. This involves breaking down the functions, such as multiplication, into several MOCs. A single MUL operation can require up to 1600ns as described in DRISA [6]. To assess the impact of such time-consuming operations on the overall transformers' computational execution time, we conducted a detailed analysis focusing on the computations performed within transformer layers in encoder-only [2], [3] and encoder-decoder [1] architectures using the DRISA accelerator [6]. Our analysis results shown in Fig. 2 indicate that over 90% of the time spent on accelerating transformer computations is required by the DRAM arrays performing the MatMul operations in the MHA and FFN layers. This motivates optimizing the MatMul operations.

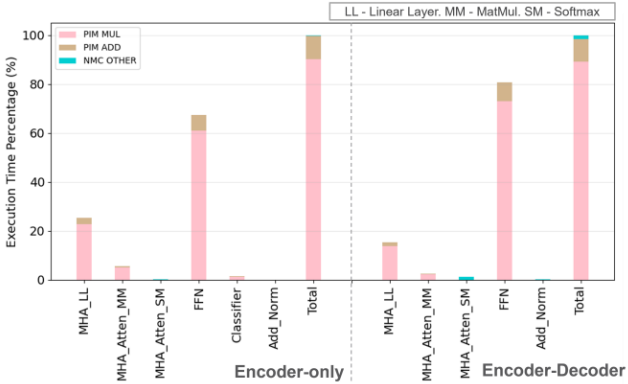

**Fig. 2.** Component-wise analysis for accelerating transformer neural network computations on traditional PIM architectures.

Prior efforts have attempted to address the MatMul bottleneck for DNN PIM acceleration. For example, a few previous works proposed using in-DRAM SC for accelerating CNNs. Such accelerators have demonstrated improvements over conventional PIM solutions. For example, SCOPE introduced a hierarchical and hybrid deterministic (H2D) SC arithmetic technique, capable of executing a single MAC operation in 200ns [13]. Another example is ATRIA which leverages bit-parallel stochastic arithmetic-based acceleration of CNNs within modified DRAM arrays that can perform 16 MACs in 85ns [19]. Other efforts explored specifically accelerating a transformer's MAC operations using alternative technologies such as ReRAM-based memory architectures, as in ReBERT [11]. However, as discussed in the previous subsection, leveraging ReRAM cells for PIM acceleration presents complex challenges. Conversely, ARTEMIS is the first accelerator to tailor in-DRAM SC specifically for transformer models. By integrating PIM and NMC, ARTEMIS employs SC for multiplication operations and analog-based computations for accumulation operations. This innovative approach significantly surpasses the computational capabilities of prior efforts, achieving 64 MAC operations in just 48 ns per subarray.

It should be noted that optimizing transformer neural network computations without sufficient optimizations for dataflow and software scheduling can still considerably limit improvements with PIM. Accordingly, ARTEMIS not only focuses on optimizing the execution of a transformer's computations but also on efficiently improving and reducing the latency involved with inter-bank and intra-bank data communication. Memory-based systems tailored for conventional DNNs usually employ optimizations in the software layer aimed at maximizing parallelism only. Accordingly, a layer-based dataflow scheme is used to allocate sufficient memory resources based on the computations in each layer. This approach necessitates loading the entire data to be processed before each layer begins executing. Previous works outlined how such approaches when extended to transformers can result in most of the execution time being spent on data handling (movement, loading, re-organization, etc.) [9]. Alternatively, employing a token-based dataflow has been proven more efficient when accelerating transformer models [9], [10]. This entails mapping the transformer computations to the memory-based system based on a token-sharding mechanism. TransPIM [9] initially introduced this approach where it implemented the token-based dataflow for transformer models. Another accelerator that elaborates on the advantages of such a scheduling approach is HAIMA [10] where a hybrid SRAM-DRAM architecture is used for the various MatMuls and data movements of their outputs. ARTEMIS adapts and enhances the token-based dataflow to our stochastic-analog computational flow for efficient inter-bank data movement while also implementing an energy-efficient intra-bank data movement micro-architecture.

### III. ARTEMIS IN-DRAM ACCELERATOR: OVERVIEW

In this section we describe our in-DRAM transformer accelerator, ARTEMIS. Within an 8GB HBM module, ARTEMIS implements minimal modifications to the conventional DRAM bank and subarray architectures, as shown in Fig. 3. In the DRAM tiles, these modifications involve incorporating small circuits (indicated in orange in Fig. 3(d)) and integrating a MOMCAP atop each tile as shown in Fig. 3(b). Additionally, within each DRAM bank, a near-subarray compute unit (NSC) is introduced for every subarray, comprising basic digital circuits and LUTs that are easily integrated and synthesized using the same DRAM memory technology at 22nm. The transformer layer operations are realized through three main computations, namely MAC, analog-to-binary conversion (A_to_B), and near-subarray computation. All modifications implemented in the DRAM tiles utilize basic digital components, such as diodes and transistors, which can be integrated through a cost-effective manufacturing process [25]. ARTEMIS follows a hardware-software co-design approach and integrates several dataflow and scheduling optimizations, allowing it to efficiently exploit the HBM's parallelism and also overcome intra-memory data movement bottlenecks. The following subsections describe the components and optimizations of our proposed architecture.

#### A. Multiply and Accumulate (MAC)
*1) Stochastic Multiplications*
While SC reduces the overall number of MOCs necessary for MAC operations during multiplications [19], it introduces considerable challenges related to output precision. Several previous SC-based accelerators for conventional neural network acceleration have attempted to tackle this issue. For example, the utilization of SCOPE's H2D SC arithmetic [13], which incorporates computational S/As, has been shown to enhance CNN inference accuracy; however, it comes with a notable increase in area overhead. ATRIA [19] addresses stochastic multiplication inaccuracies by increasing the bit width required for stochastic representation, at the expense of reducing parallelism. Another approach in [31] redesigns the stochastic multiplier to utilize transition-coded unary (TCU) numbers for realizing bit-parallel deterministic stochastic multiplications, resulting in a reduction of computational errors by up to 32.2%. However, the implementation in [31] requires the integration of additional circuits and logic gate arrays.

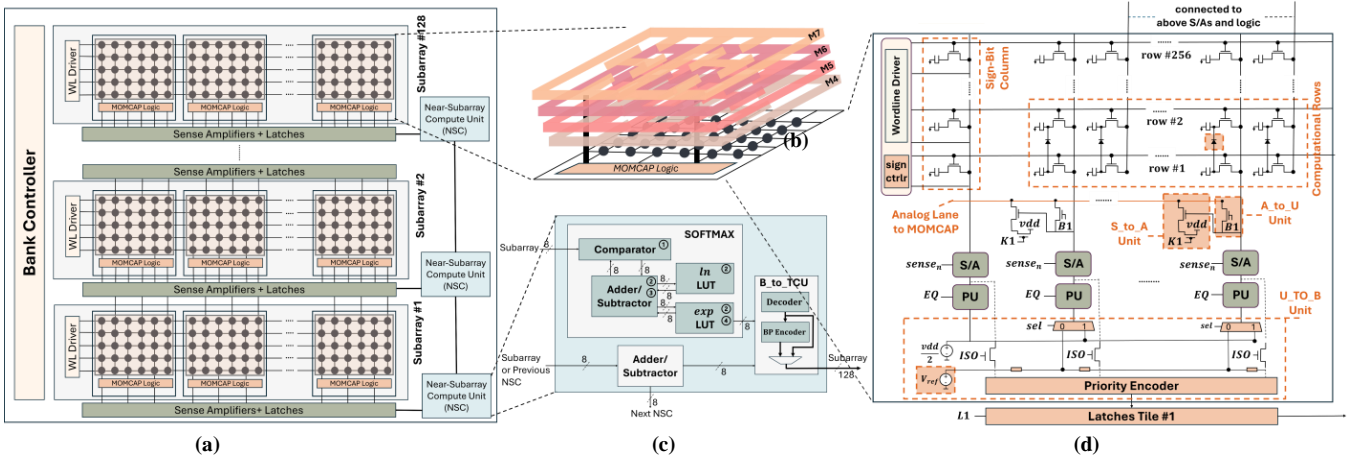

**Fig. 3.** ARTEMIS architecture overview showing (a) design of a single bank composed of **128** subarrays, each with **32** tiles, (b) schematic layout of MOMCAP using metal layers (M4-M7), (c) structure of the first NSC unit, (d) structure of the first tile.

In contrast to relying on a multiplier circuit like the one described in [31], ARTEMIS introduces deterministic stochastic multiplication utilizing TCU numbers within the DRAM bit-line logic. TCU numbers are stochastic bit-streams where all the '1's are grouped at either of the stream's trailing ends. This approach eliminates the need for additional circuitry within DRAM tiles, enabling the exploitation of parallelism while minimizing area overhead and mitigating SC multiplication inaccuracies.

Initially, the transformer layer parameters are distributed across ARTEMIS subarrays. When performing multiplications, to ensure accurate operation of the deterministic multiplication method, the first operand is generated using a binary-to-transition-coded-unary (B_to_TCU) decoder, followed by a bit-position correlation encoder, while the second operand is generated using a B_to_TCU decoder only. Each multiplication operation involved in the MatMuls in a transformer's MHA and FF layers is then performed stochastically.

In contrast to previous stochastic in-DRAM transformer accelerators, which require multiple MOCs or complex multiplier circuits [13], ARTEMIS computes one multiplication operation by executing only two MOCs to copy the operands into two distinct computational rows. This is achieved by extending the method in [30] for fast and energy-efficient SC logic operations where ARTEMIS reserves the entire first two rows in each subarray for SC multiplications. As shown in Fig. 3(d), these two rows are connected with diodes between each pair of bit-cells and the AND result is thus computed and stored in the first computational row. A read operation is subsequently performed by pre-charging the bit-lines using the EQ signal which controls the pre-charge unit (PU). Computational row #1 is then activated by asserting $WL_{comp\_row1}$, and enabling the S/As using the $sense_n$ signal.

Our baseline memory architecture incorporates an open-bit-line approach [12] where only half of each DRAM bank's subarrays are operated concurrently at a time. Thus, as shown in Fig. 3(a), each DRAM tile is connected to two sets of S/As, where one half of the bit-lines (128 out of 256 columns) are operated using the S/As set at the bottom, while the other half are connected to the set at the top. ARTEMIS represents signed 8-bit binary numbers as 128-bits stochastic streams plus 1 sign bit, which is captured using a per-subarray added bit-line column indicating the sign associated with the numbers stored in each row. Accordingly, each row in a tile stores all positive or all negative numbers and each tile can process up to two multiply operations at a time.

*2) Analog Temporal Accumulations*

Stochastic-based addition has been shown to introduce considerable errors [13]. In pursuit of both accuracy and speed during addition operations, we utilize analog accumulation facilitated by a MOMCAP within each DRAM tile in the HBM. ARTEMIS repurposes S/As to convert the number of 1's in a stochastic product value into a proportional analog voltage on the MOMCAP. This serves to convert the stochastic product value into an analog representation. Multiple analog voltage values representing multiple different stochastic product values can be sequentially accrued on the MOMCAP via analog accumulation. The customized H-shaped MOMCAP, shown in Fig. 3(b), optimizes capacitance without increasing the overall tile area of ARTEMIS. They are integrated into DRAM arrays with minimal modification to the array itself, since they are implemented using different metal layers stacked on DRAM tiles. The feasibility of incorporating MOMCAPs within DRAM structures has been effectively demonstrated in [32]. Recent advancements in VLSI enable the seamless incorporation of MOMCAP capacitors using While prior research, such as [32] replaced conventional embedded-DRAM cell capacitors with similar MOMCAPs to extend retention times, ARTEMIS is the first in-DRAM design to incorporate MOMCAPs for in-DRAM analog computing purposes.

The capacitance of the MOMCAP is contingent upon the capacitor's area, which determines the maximum number of consecutive accumulations it can accommodate. A higher number of accumulations enhances performance by reducing the need for frequent data conversions. However, as MOMCAPs are constructed using metal layers (M4-M7), their area must align with that of the tile to prevent an increase in overall size. Thus, we conducted a detailed analysis to determine the maximum number of accumulations achievable with varying capacitance values. An appropriate area budget to support up to 20 consecutive accumulations for each MOMCAP was thus established (see results in Section IV.B).

Each MOMCAP is connected to an analog lane which is connected directly to the S/A circuits, as shown in Fig. 3(d). To enhance performance and achieve higher parallelism, each operational DRAM tile performing two multiplications at a time utilizes two MOMCAPs; its own as well as that of the non-operational DRAM tile above or below it as shown in Fig. 4. Accordingly, up to 40 MAC operations can be accommodated by each operational DRAM tile before requiring any data movement or conversions. The accumulation operation proceeds as follows: following one multiplication operation and storage of the output bits by the tile's S/As, each bit-line holds a value of '1' or '0'. To convert this stochastic data into analog charge for accumulation on the MOMCAP, a stochastic-to-analog (S_to_A) circuit is implemented, comprising two transistors (Fig. 3(d)). This configuration supplies adequate voltage for the capacitor to detect all necessary voltage level changes. Upon toggling signal $K_1$, all bit-lines within the same tile connect to the two MOMCAPs (Fig. 4), resulting in two concurrent accumulations of charge, each directly proportional to the number of its connected bit-lines storing '1' values. Subsequently, as the following sets of operands undergo multiplication, their two outputs are once again stored in the two MOMCAPs, effectively adding to the previous multiplication results.

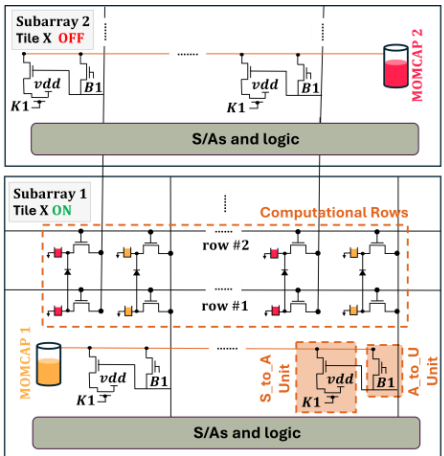

**Fig. 4.** MOMCAPs charging during analog accumulation step.

### B. Analog to Binary Data Conversion

The analog values preserved within each tile's MOMCAP require conversion into binary numbers for subsequent processing upon reaching the MOMCAP's charge capacity. ARTEMIS refines the circuits and timing signals from AGNI [18], achieving a reduced latency of 31ns for the S_to_B conversion compared to AGNI's 56ns. The enhanced S_to_A conversion circuit is described in the previous subsection. ARTEMIS employs a two-step process for analog-to-binary conversion: analog-to-transition-coded-unary (A_to_U) and transition-coded-unary-to-binary (U_to_B). Activation of the A_to_U circuit involves toggling control signal $B_1$ to connect the stored MOMCAP value and the tiles' bit-lines. Subsequently, the S/As are repurposed as voltage comparators by pre-charging bit-lines to distinct voltage levels determined by the voltage divider circuit. The MUX *sel* signal controls the voltage divider circuit. This process yields A_to_U data conversion. Next, activation of the U_to_B unit is initiated by asserting the *ISO* signal, allowing the TCU number to traverse a priority encoder. Finally, each tile's binary result is latched for transmission to an NSC unit (discussed in subsection III.C).

### C. Near-Subarray Compute Unit (NSC)

The NSC unit is composed of simple digital circuits and LUTs with one NSC assigned to each subarray. It handles the acceleration of the tiles' partial sum accumulations, non-linear functions, and B_to_TCU data conversions.

#### 1) Reduction Operations
Following the computation of 40 MAC operations as explained in the previous sections, each tile in the bank will have a partial sum output stored in its local latches. All the tiles' partial sums need to thus be gathered and reduced. Each subarray's NSC unit is equipped with a 2-input 8-bit binary adder/subtractor to handle the partial sum accumulations. Section III.D.2 outlines the intra-bank data movement scheme applied in ARTEMIS to efficiently handle transferring all the tiles' data to the NSC units. Each subarray's NSC is responsible for accumulating all the partial sums computed in that subarray. Additionally, each NSC manages the accumulation of the output from the NSC unit following it, as illustrated in Fig. 5(a). In the example used in the figure, NSC 1 and NSC 2 first accumulate all the values output from their respective subarrays in sub-round 2. Afterwards, NSC 1 receives and accumulates the resultant output from NSC 2 in sub-round 3. To accommodate both positive and negative numbers, ARTEMIS performs MAC operations initially for all positive numbers (identified by the sign-bit column), consolidating the final positive result at each subarray's NSC unit. This process is then repeated for negative numbers, with their result subsequently subtracted from the positive result previously gathered using the same adder/subtractor block in each NSC.

#### 2) Softmax
Each NSC unit is equipped with reprogrammable LUTs to handle fast execution of non-linear functions. Non-linear functions such as ReLU (used in FFN layers) and GELU (used in ViTs) can be realized using stand-alone LUTs. However, the softmax function that is frequently required in each head of the MHA layers, poses two main challenges. First, as expressed in Eq. (5) below, softmax involves computationally expensive division and numerical overflow operations. Second, exploiting parallelism is a non-trivial task since all results from the previous MatMul need to be generated first before computing the softmax output for each value. To overcome both challenges, we employ the log-sum-exp approach, used in various previous works such as [33] as shown in the equation:

$$Softmax(y_i) = \frac{\exp(y_i - y_{max})}{\sum_{j=1}^{D} \exp(y_j - y_{max})},$$
$$= \exp\left(y_i - y_{max} - \ln\left(\sum_{j=1}^{D} \exp(y_j - y_{max})\right)\right), \quad (5)$$

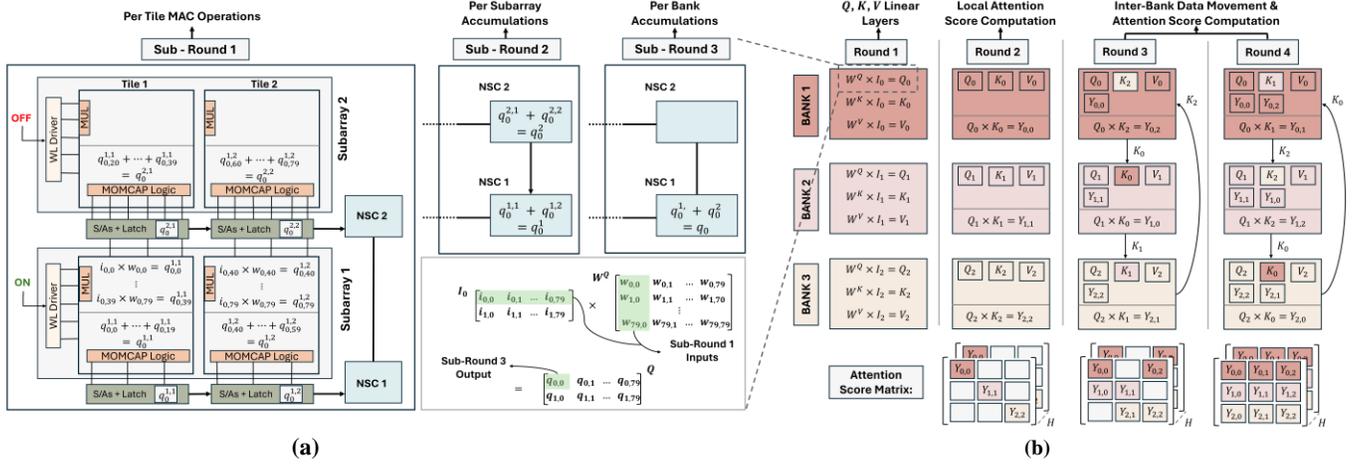

**Fig. 5.** ARTEMIS dataflow scheme examples showing: (a) per-subarray vector multiplication flow with 2 subarrays and 2 tiles, (b) token-based dataflow scheme for computing attention scores in MHA layers with 3 banks.

This allows us to divide the softmax execution into four main operations: ① finding $y_{max}$; ② performing $ln(\sum_{j=1}^{D} exp(y_j - y_{max}))$; ③ subtracting $(ln)$ output from $(y_i - y_{max})$, and; ④ performing the final $(exp)$ function. As the $Y$ matrix is being generated from the MatMul preceding the softmax operation $(QK^T)$ in the scaled dot product attention block, the output $y_i$ is fed directly to a 2-input 8-bit comparator with a local register to hold the current $y_{max}$, thus pipelining the execution of ①. Following the generation of matrix $Y$ and storing $y_{max}$ in all NSC units, ② is computed using the blocks labelled with "②" in Fig. 3(c). Subtraction ③ is then performed using the softmax adder/subtractor and finally, ④ is computed using the $exp$ LUT. The orchestration of data movement and pipelining of softmax is further elaborated on in section III.D.2.

*3) Binary to TCU Data Conversion*

The transformer's intermediate results are inputs to the next operations or layers. For example, the softmax output $S$ in the MHA's scaled dot-product attention evaluation, is used to compute $S \times V$ (see Fig. 1). Accordingly, all values in matrix $S$ need to be converted from binary to stochastic bitstreams to be used in stochastic multiplications. As explained in section III.A.1, ARTEMIS uses a deterministic multiplication method, where the first operand is generated using a B_to_TCU decoder, followed by a bit-position correlation encoder, while the second operand is generated using a B_to_TCU decoder only. Thus, the B_to_TCU block in each NSC unit comprises of a B_to_TCU decoder and a bit-position correlation encoder as shown in Fig. 3(c). Depending on the order of the operand, the output of the B_to_TCU block will be that of the B_to_TCU decoder only or that of the bit-position correlation encoder. The bit-position correlation encoder ensures that the conditional probability of the 1st operand given the 2nd operand matches the marginal probability of the 1st operand [18].

*D. Dataflow and Scheduling Optimizations*

*1) Dataflow and Inter-bank Communication*

To maximize HBM parallelism and overcome the data movement bottleneck when accelerating transformer models with a layer-based dataflow [6], [34]-[36], ARTEMIS adapts a token-based data sharding dataflow [9], modified for its stochastic-analog computational flow.

In a transformer model, a sequence input is initially transformed into a series of input embeddings, where each embedding vector corresponds to a 'token' [1]. Each token encapsulates specific features associated with the input sequence. Layer-based dataflow maps all the tokens to the same bank(s) responsible for computing the first transformer layer. All data output from the first layer is then transferred to the next bank(s) associated with performing the next layer's computations. Given the large number of model parameters in a transformer and the shared data bus of HBM, which allows only one bank to transfer its data at a time [12], this leads to significantly high congestion and data movement latencies.

Alternatively, token-based dataflows map the data across the HBM banks based on input tokens. The primary advantage of employing token-based data sharding is the facilitation of data reuse across various layers by consolidating computations of tokens within the same memory location. This approach reduces the cost of data movement while capitalizing on memory-level parallelism, as different banks can independently handle computations and data movements for allocated tokens.

Following token sharding, each bank manages computations for its assigned segments throughout the entire transformer inference process. Token-based data sharding is implemented on input tokens before the linear layers of the initial encoder block. Accordingly, when the number of tokens, $N$, used in a model is greater than the number of banks, $K$, in the HBM module, each bank will operate on $N_b = \frac{N}{K}$ number of tokens.

To exploit the parallelism and performance improvements offered by our architecture's stochastic-analog computational scheme, ARTEMIS utilizes each tiles' row of latches and the NSCs to handle data being placed on or received from the HBM's links. Prior to transferring the banks' data to its neighboring bank, the stochastic output is converted to binary using the per-tile B_to_S circuits, which significantly reduces the number of bits transferred. Upon arrival to the neighboring bank, the data is first received by the NSC units where it is input to the B_to_S block. Using the per-tile latches rows, the stochastic numbers are then moved in a pipelined manner to the appropriate tiles where they are directly written to the target and computational rows to be used in the next computations.

Fig. 5(b) illustrates an example of processing the first linear layers ($Q, K, V$ $generation$), and the attention score computation ($Y = QK^T$) in the MHA layer. Initially the input matrix is distributed based on the token-sharding mechanism explained above, where each bank will operate on ($I_i \in R^{N_B \times D}$). In Round 1, each bank will generate its own local $Q_i, K_i$, and $V_i$, each with size $N_B \times D$. Each bank then computes its local attention scores using the stored $Q_i$ and $K_i$, and by the end of Round 2, each bank will have generated the partial attention score matrix $Y_{i,i}$. To correctly generate the complete attention score matrix, each bank will need to transfer its own $K_i$ matrix to all other banks. Similar to TransPIM [9], a ring and broadcast network is utilized to minimize the latency cost of the data movement steps in Rounds 3 and 4. As each bank $i$ receives the partial $K_j$ matrices from all the banks, it will keep on generating partial attention score matrices $Y_{i,j}$ till all the values are computed in Round 3. The next steps in the MHA layer entail the softmax operation and the attention output computation (($S_i \times V_i$). When performing the latter, rounds 2, 3 and 4 will need to be repeated as partial $V_i$ will also need to be exchanged between all the banks for correct operation.

*2) Intra-bank Communication*

Fig. 5(a) outlines the underlying operation flow in the bank 1 subarrays when generating one value in the $Q$ matrix. In this example the dimension of $Q$ is 80 and thus to calculate the first value, $q_{0,0}$, the first row from the partial input matrix $I_0$ needs to be multiplied by the first column in the query weight matrix $W^Q$. This results in vector multiplication with size 80.

As explained in section III.A, ARTEMIS follows an open-bit-line architecture where only half the subarrays in a bank are activated at a time. Accordingly, in the example in Fig. 5(a), only one out of the two subarrays will be activated concurrently. For simplicity, we also assume that only subarray 1 is "ON" for all the vector multiplication operations. As discussed in sections III.A and III.C, each tile can perform 40 MAC operations before converting the accumulated analog value stored in the MOMCAPs to binary values. Thus, tile 1 in subarray 1 will perform stochastic multiply operations using sub-vectors $I_0[0:39]$ and $W^Q[0:39]$ and perform the analog temporal accumulations for multiply outputs 0 to 19 only. Meanwhile, tile 1 in subarray 2 will accumulate multiply outputs 20 to 39 using its own MOMCAP and associated logic. Similar operations will be computed in tiles 2 in subarrays 1 and 2.

By the end of Sub-Round 1, each tile's binary partial sum output will be stored in the tile latches. These values will then be transferred to the NSC units in a pipelined manner, till both values from each subarray reach the NSC and are immediately added using the adder/subtractor circuit as shown in Sub-Round 2. The last step (Sub-Round 3) is then to move the partial sum output from NSC 2 to NSC 1 to be further reduced into $q_{0,0}$. Since the sign bits column corresponds to both values stored in each operational tile, in this example, NSC 1 is responsible for forwarding the sign bit to NSC 2 as well.

*3) Execution Pipelining and Scheduling*

To further exploit parallelism, ARTEMIS pipelines the transformers' operations. Fig. 6 outlines the pipelining model adopted by our architecture when accelerating an MHA layer in one bank. The MHA operations are divided into 8 steps as shown in the top half of Fig. 6. First, when generating the $Q_i, K_i$, and $V_i$ matrices, ARTEMIS pipelines the following: (i) performing the in-situ MAC operations within the DRAM tiles, (ii) pipelining the data movement using the row of latches and (iii) accumulating the binary partial sums in the NSC units. As shown in Fig. 6, this efficiently hides the latencies associated with the intra-bank data movement and the NSC reduction operations. This pipelining scheme is applied when performing any MatMul operations in the MHA and FFN layers in the transformer's encoder or decoder blocks. After generating the local attention score partial matrix by computing $Q_i \times K_i^T$, each bank will need to send its local $K_i$ matrix to all other banks using the ring and broadcast technique discussed earlier.

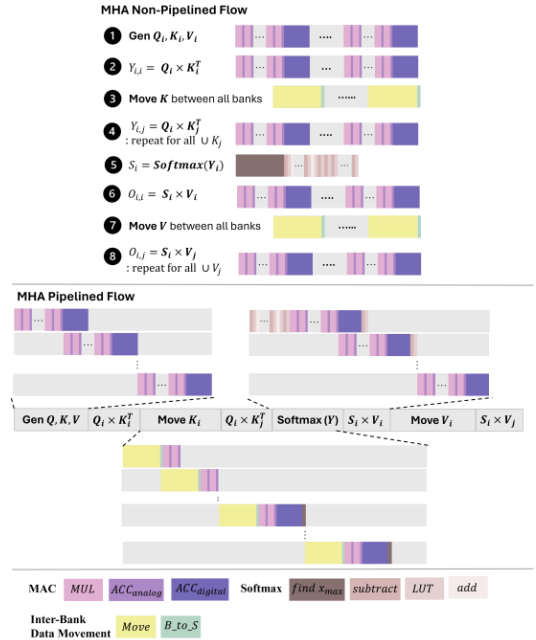

**Fig. 6.** ARTEMIS pipelining within one bank for MHA layers.

While ARTEMIS significantly reduces the latencies associated with performing transformer operations, the inter-bank data movement step is predominately the most time-consuming step based on our analysis. Nevertheless, our hardware accelerator mitigates the latency of this step by overlapping the inter-bank data movement with the B_to_S data conversions, softmax, and the next MatMul to be executed ($S_i \times V_i$) as shown in the pipelined flow in Fig. 6. Data is transferred between banks in binary using a 256-bit link and as new data arrives to a bank, instead of first writing the value to the DRAM arrays, ARTEMIS directly passes it through the B_to_TCU blocks in the NSC units to prepare the stochastic multiplication operands. These values are then written in the tiles' computation rows to be used immediately in the MAC operations. Such optimizations not only result in faster execution but also reduce energy consumption associated with the eliminated DRAM write operations. As the attention score matrices are being generated in each bank, the output values are being input concurrently to the softmax 8-bit comparators to keep updating $y_{max}$ (see Eq. (5)). Other softmax operations such as the subtractions and the final exponent calculation are also pipelined when computing $(S_i \times V_i)$ as shown in Fig. 6.

## IV. EXPERIMENTAL RESULTS

We developed a comprehensive simulator in Python to estimate the performance and energy costs of our proposed accelerator by accurately modeling all hardware components and in-DRAM operations. The simulator considers both software and hardware mapping, while performing the layer-wise mapping for each transformer model and dataset. The costs associated with each modeled hardware component were derived through extensive analysis and simulations. DRAM area estimates were obtained using CACTI-3D [37], while latency values for per-tile circuits were calculated using detailed LTSPICE simulations. All circuits within the NSC units and latches were synthesized using Cadence Genus, with the resulting latency, power, and area values reported in Table III. Lastly, the energy values for HBM operations are based on specifications from Samsung's HBM [12], as shown in Table I, based on 22nm DRAM technology. $e_{act}$ is the activation energy associated with an ACTIVATE operation for a DRAM row in one bank. The datapath energies for moving data within the DRAM chips are composed of 1) traversing the local data-lines and the master data-lines from the row buffer to the global S/As (GSA) ($e_{Pre-GSA}$), 2) traversing the path from the GSAs to the DRAM I/Os ($e_{Post-GSA}$), and 3) traversing the I/O channel between the DRAM and GPU ($e_{I/O}$) [12].

The DRAM bank structure in our architecture is slightly re-arranged in comparison to previous work and conventional HBM architectures [9], [12]. Each subarray is comprised of only 256 rows, allowing for faster operation per subarray and higher parallelism. While this results in slightly increased area and power consumption, such organization is better aligned with SC. Based on our SPICE simulations, one MOC in ARTEMIS is equivalent to $17ns$. Moreover, the overall power budget for ARTEMIS is $60W$, in alignment with the HBM conventional DRAM power budget [12]. Five transformer model workloads were considered in all our experiments: Transformer-base, BERT-base, ALBERT-base, ViT-base, and OPT-350. Details of these models are shown in Table II.

TABLE I
ARTEMIS HBM CONFIGURATION PARAMETERS

| | Parameters | Value |
|---|---|---|
| **Configuration** | Number of HBM stacks | 1 |
| | Number of channels per stack | 8 |
| | Number of banks per channel | 4 |
| | Number of subarrays per bank | 128 |
| | Number of tiles per subarray | 32 |
| | Number of rows per tile | 256 |
| | Number of bits per row | 256 |
| **Energy** | $e_{act}$= 909 pJ, $e_{Pre-GSA}$= 1.51 pJ/b, $e_{Post-GSA}$=1.17/b, $e_{I/O}$= 0.80 pJ/b | |

TABLE II
TRANSFORMER MODEL CONFIGURATIONS

| Model | Params | Layers | N | Heads | $d_{model}$ | $d_{ff}$ |
|---|---|---|---|---|---|---|
| **Transformer-base** | 52M | 2 | 128 | 8 | 512 | 2048 |
| **BERT-base** | 108M | 12 | 128 | 12 | 768 | 3072 |
| **Albert-base** | 12M | 12 | 128 | 12 | 768 | 3072 |
| **ViT-base** | 86M | 12 | 256 | 12 | 768 | 3072 |
| **OPT-350** | 350M | 12 | 2048 | 12 | 768 | 3072 |

TABLE III
ARTEMIS PER SUBARRAY HARDWARE OVERHEAD

| Component | Latency ($ps$) | Power ($mW$) | Area ($\mu m^2$) |
|---|---|---|---|
| **S_to_B Circuits** | 20000 | 0.053 | 970 |
| **Comparator** | 623.7 | 0.055 | 0.0088 |
| **Adder/Subtractors** | 719.95 | 0.0028 | 0.0055 |
| **LUTs** | 222.5 | 4.21 | 4.79 |
| **B_to_TCU Blocks** | 530.2 | 0.021 | 0.063 |
| **Latches** | 77.7 | 0.028 | 0.13 |

### A. Computational Error and Accuracy Analysis

Given that SC demands $2^N$ bits for each N-bit binary number, neural network model compression, particularly through quantization, can enhance the overall performance. Our analysis indicates that the utilization of 8-bit model quantization results in transformer inference accuracy levels comparable to those achieved with full precision (FP32), as depicted in Table IV. The % accuracy metric is used to assess transformer-base, BERT-base, Albert-base, and ViT models that are used for translation, sentiment analysis and image classification tasks respectively. Meanwhile the BLEU score metric is reported for the OPT-350 model that is used for a text-generation task. Based on this analysis, we have opted for transformer models featuring 8-bit precision, where ARTEMIS represents parameter values stochastically with 128 bits plus one sign bit. Furthermore, we conducted detailed error analysis to assess the efficacy of each approximate computing operation in ARTEMIS as shown in Table V. The calibration accuracy represents the threshold in bits below which the computation results remain entirely accurate. For instance, in the case of stochastic multiplication, the output will begin to show small errors when the binary numbers involved exceed 4.68 bits in length. The mean absolute errors (MAEs) normalized to the maximum voltage supported by each operation, were accumulated and integrated into each transformer model inference. The resultant accuracy drop was found to be minimal as shown in table IV.

TABLE IV
TRANSFORMER MODEL METRICS

| Model (metric) | Dataset | FP32 | Q(8-bit) | Q(8-bit) + SC |
|---|---|---|---|---|
| **Transformer-base** | Ted-hrlr | 70.90% | 70.40% | 69.45% |
| **BERT-base** | GLUE | 87.00% | 86.27% | 85.92% |
| **Albert-base** | GLUE | 86.07% | 84.80% | 84.51% |
| **ViT-base** | ImageNet | 97.60% | 96.50% | 96.20% |
| **OPT-350** | Openassistant-Guanaco | 18.07 (BLEU) | 17.79 (BLEU) | 17.49 (BLEU) |

TABLE V
ARTEMIS PER-COMPONENT CALIBRATION ACCURACY

| Block | MAE | Max Error | Calibration Accuracy |
|---|---|---|---|
| **Stochastic MUL** | 0.039 | 0.123 | 4.68 |
| **Analog ACC** | 0.0085 | 0.0729 | 6.88 |
| **A_to_B** | 0.00037 | 0.00062 | 11.38 |
| **Softmax** | 0.0020 | 0.0078 | 8.20 |

Table IV presents the inference accuracies for the models employed in our experiments, for the baseline FP32, quantized 8-bit precision, and quantized 8-bit precision with SC multiplications cases. Through the avoidance of stochastic additions and the adoption of an optimized approach to stochastic multiplications, ARTEMIS demonstrates minimal accuracy degradation, averaging at 1.4% compared to FP32 and 0.5% compared to quantized 8-bit models.

### B. MOM Analog Capacitor Accumulation Analysis

To determine the optimal parameters for our custom MOMCAP within the DRAM tiles, we carefully modeled and simulated 128 bit-lines alongside the tile's circuits (shown in Fig. 3(d)) utilizing LTSPICE. Through this process, we analyzed the voltage behaviour of charge accumulation on the MOMCAP across a spectrum of capacitance values, ranging from $4pF$ to $40pF$, which are distinguished by various colors in Fig. 7. The linearity and symmetry observed in the steps of charge accumulation on the MOMCAP denote its stable performance and its ability to accurately differentiate between distinct voltage levels [38]. Based on our detailed experimental and numerical analysis, such behavior was a result of accurately controlling the charging time of each step, which was set to $1ns$. Each voltage increment in the graph represents the accumulation of a 128-bit number. Consequently, the maximum number of accumulations corresponds to the number of linearly increasing voltage steps until saturation occurs.

As depicted in Fig. 7, increased capacitance enhances the capacitor's ability to accommodate a greater number of accumulations. Nonetheless, as previously outlined, higher capacitance leads to a larger area overhead. Hence, we have opted for a MOMCAP size aligning with ARTEMIS' tile area of $338\mu m2$, which corresponds to an $8pF$ capacitance. This selection enables the accumulation of 20 consecutive dot products per MOMCAP.

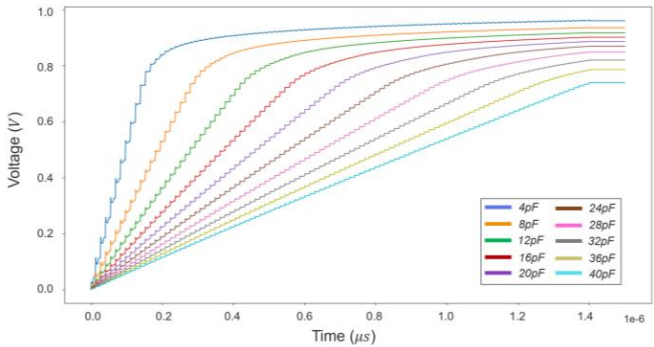

**Fig. 7.** ARTEMIS experimental results for MOMCAP voltage behavior when storing multiple consecutive accumulations of 128-bit numbers from the DRAM tile bit-lines.

### C. Dataflow and Scheduling Optimization Analysis

We conducted a sensitivity analysis to assess the impact of the dataflow and execution pipeling optimizations described in Section III.D. The speedup and normalized energy results are shown in Fig. 8(a) and 8(b), respectively. The results were obtained for executing the five transformer models on ARTEMIS but using a layer-based dataflow scheme without pipelining (layer_NP), a layer-based dataflow with pipelining enabled (layer_PP), a token-based dataflow without pipelining (token_NP), and finally our main ARTEMIS architecture with token-based dataflow and execution pipelining (token_PP).

Despite HBM offering a bandwidth of up to 256GB/s per stack, the shared data link and the massive amount of values that needs to be moved between the different transformer layers vastly limit the acceleration of transformers on PIM systems. On the other hand, utilizing the token-based data sharding dataflow explained in Section III.D.1, results in an average speedup of 11.0× without pipelining enabled and 10.8× when pipelining is enabled in both dataflow schemes. As shown in Fig 8(b), employing the token-based dataflow is also more energy efficient since the amount of data movement is reduced. An average energy reduction of 3.5× is observed without pipelining and also with execution pipelining enabled. Pipelining also has an impact on speedup and energy since ARTEMIS efficiently pipelines various operations within each layer. The energy reduction is also due to avoiding unnecessary write operations when receiving new data from neighboring banks. On average, pipelining results in a speedup of 50% with the layer-based dataflow and 43% with the token-based dataflow. For energy consumption, pipelining results in 42% energy reduction with the layer-based dataflow and 43% reduction with token-based dataflow. We observed that the impact of pieplining and the token-based dataflow was greatest when accelerating ViTs. This is partly due to the longer input sequences that still fit onto our architecture, used with the ViT model in our experiments. Meanwhile, OPT exhibited slightly lower speedups since its sequence length is larger than the total number of banks in the ARTEMIS baseline hardware configuration. This however indicates promising scalability results which is further elaborated on in section IV.E.

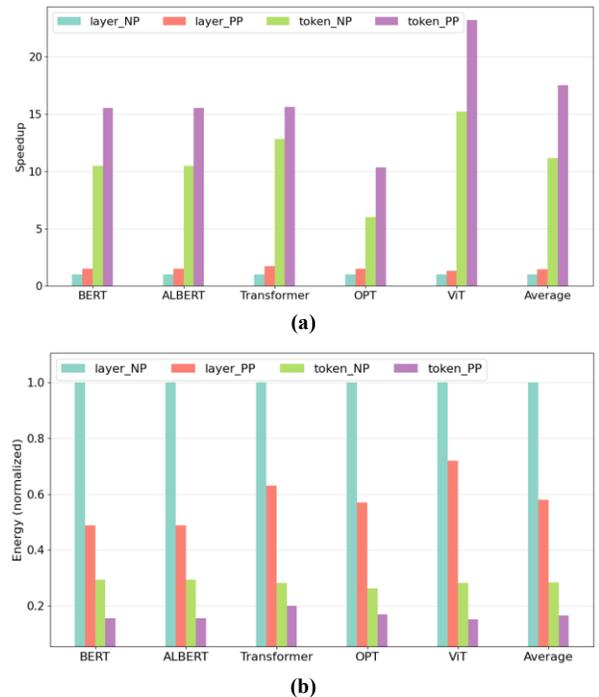

**Fig. 8.** Sensitivity analysis showing the impact of token-based dataflow and execution pipelining on (a) speedup, (b) energy.

### D. Comparison with State-of-the-art Computation Platforms

We compared ARTEMIS with CPU, GPU, TPU, several state-of-the-art PIM transformer accelerators: TransPIM [9],

HAIMA [10], and ReBERT [11], and an FPGA-based transformer accelerator (FPGA_ACC) [40]. Note that ReBERT only focuses on BERT-based models and is not included in the comparisons for the other models. We used power, latency, and energy values reported for the selected accelerators, and directly obtained results from executing models on the GPU, CPU, and TPU platforms to estimate the energy, power efficiency, and inference latency for each model and dataset.

*1) Speedup Comparison*

Fig. 9 shows the speedup comparison between ARTEMIS, the compute platforms, and the transformer PIM accelerators considered. The speedup values are all relative to the CPU inference latency. On average, ARTEMIS achieves 1230×, 157×, 212×, 29.6×, 4.8×, 11.9×, and 3.6× speedup compared to CPU, GPU, TPU, FPGA_ACC, TransPIM, ReBERT, and HAIMA, respectively. The lower latencies observed with ARTEMIS can be attributed to its ability to perform 64 MAC operations in only $48ns$ using SC and analog-based computing. Furthermore, our optimized data mapping, movement, and scheduling schemes aided in reducing the overall latency.

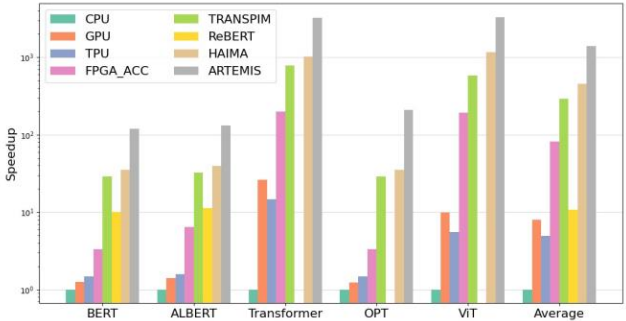

**Fig. 9.** Speedup comparison between ARTEMIS, CPU, GPU, TPU and PIM accelerators.

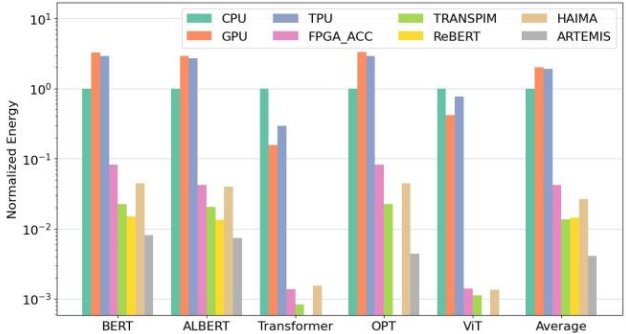

**Fig. 10.** Energy comparison between ARTEMIS, CPU, GPU, TPU and PIM accelerators.

*2) Energy Comparison*

The energy comparison results for ARTEMIS with the computing platforms and transformer PIM accelerators considered are shown in Fig. 10. All the energy values are normalized to the CPU. ARTEMIS achieved on average 1443.3×, 700.4×, 1000.4×, 8.8×, 3.5×, 1.8×, and 6.2× lower energy values compared to CPU, GPU, TPU, FPGA_ACC, TransPIM, ReBERT, and HAIMA, respectively. The reduced energy consumption observed with our architecture can be explained in terms of the significantly reduced number of required DRAM row activations when accelerating transformers' predominant computations, namely MACs. This results from SC enabling the compute-intensive multiplication operations to be realized using simple in-DRAM AND operations along with the MOMCAP analog compute logic facilitating fast and energy-efficient analog accumulations.

*3) Power Efficiency Comparison*

Fig. 11 shows the power efficiency results (in terms of GOPS/Watt values) when comparing ARTEMIS to all other compute platforms and PIM accelerators. Our accelerator attains on average 1269.0×, 673.6×, 950.2×, 8.5×, 3.3×, 1.9×, and 5.9× improvement compared to CPU, GPU, TPU, FPGA_ACC, TransPIM, ReBERT, and HAIMA, respectively. The enhanced power efficiency of ARTEMIS is due to its notable low per-MAC latency, and the overall high throughput operation while abiding by a maximum power budget of 60W. Moreover, by employing the various compute, data movement, and orchestration optimizations explained earlier, our architecture can efficiently accommodate all the various transformer models' operations using minimal added circuitry. This in turn, significantly improves the power efficiency.

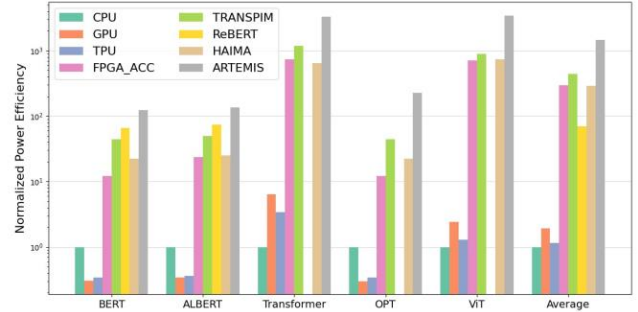

**Fig. 11.** Power efficiency ($GOPS/W$) comparison between ARTEMIS, CPU, GPU, TPU and PIM accelerators

### E. Scalability Analysis

Transformer models usually encounter considerable challenges when handling long input sequences [39]. Conventional platforms such as CPUs, GPUs, and TPUs are constrained by the sequence length due to their limited available memory capacity. Meanwhile, PIM-based systems present a promising avenue for scalability, offering the potential for enhanced memory bandwidth while concurrently increasing parallelism with minimal memory access latency. Illustrated in Fig. 12 are the speedup outcomes obtained by employing additional HBM stacks for processing workloads of increasing input sequence lengths. It is evident that larger hardware configurations, which provide a greater number of banks, yield increased speedups with longer sequence lengths. This is because more token groups can fit onto the accelerator, minimizing the need for multiple mappings and the associated latency overhead. Overall, the speedup results averaged across all transformer models used, demonstrate that ARTEMIS exhibits commendable scalability, approaching near-linear performance enhancement for extended sequence workloads that fully utilize the computational capabilities of HBM. Although power consumption can increase with more HBM stacks, the substantial speedup achieved ensures that energy efficiency is still enhanced. Employing larger hardware sizes for larger models circumvents the additional energy expenditure associated with repeatedly writing and mapping the models' parameters to the DRAM banks when the models do not fit on the accelerator. These and previous experimental findings strongly suggest that combining concepts of stochastic and analog computing in PIM systems while utilizing optimized dataflow schemes enable a viable and efficient solution for accelerating long-sequence transformer applications.

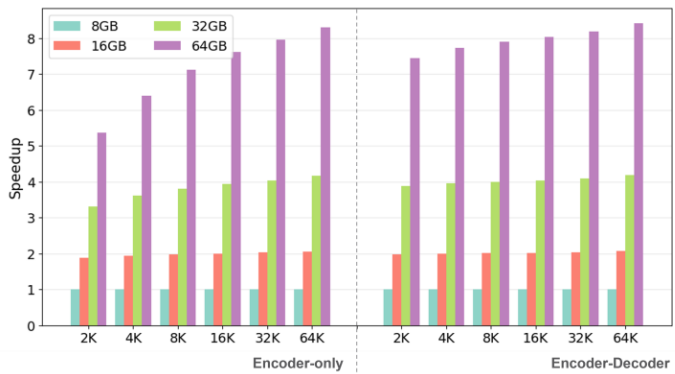

**Fig. 12.** ARTEMIS scalability analysis when increasing the input sequence length for transformer neural network models.

## V. CONCLUSION

In this article, we presented a novel in-DRAM hardware accelerator for transformer neural networks that combines stochastic and analog computing and extends state-of-the-art HBM architectures. Our proposed ARTEMIS architecture demonstrated remarkably low per-MAC latency through the utilization of bit-parallel stochastic computing for multiplications, coupled with analog domain accumulations. ARTEMIS exhibited at least 3.0× speedup, 1.8× lower energy, and 1.9× better power efficiency when compared to GPU, TPU, CPU and multiple state-of-the-art PIM transformer accelerators. The results demonstrate the promise of utilizing in-DRAM stochastic and analog computations for transformer neural network acceleration.

## REFERENCES

[1] A. Vaswani, N. Shazeer, N. Parmar, J. Uszkoreit, L. Jones, A. N. Gomez, L. Kaiser, and I. Polosukhin, "Attention is all you need," *NIPS*, 2017.
[2] J. Devlin, M. Chang, K. Lee, and K. Toutanova, "BERT: pre-training of deep bidirectional transformers for language understanding," *CoRR*, Oct 2018.
[3] Z. Lan, *et al.,* "Albert: A lite bert for self-supervised learning of language representations," *ICLR*, Sep. 2019.
[4] J. Achiam, *et al.,* "Gpt-4 technical report," *arXiv preprint arXiv:2303.08774*., 2023.
[5] A. Dosovitskiy, *et al.*, "An image is worth 16x16 words: Transformers for image recognition at scale," *ICLR*, Oct. 2020.
[6] S. Li, *et al.,* "Drisa: A dram-based reconfigurable in-situ accelerator," *IEEE/ACM MICRO*, pp. 288-301, 2017.
[7] S. Lu, M. Wang, S. Liang, J. Lin, J. and Wang, Z. , "Hardware accelerator for multi-head attention and position-wise feed-forward in the transformer," *IEEE SOCC*, 2020.
[8] Q. Panjie, *et al*., "Accelerating framework of transformer by hardware design and model compression co-optimization," *IEEE ICCAD*, 2021.
[9] M. Zhou, W. Xu, J. Kang and T. Rosing, "TransPIM: A Memory-based Acceleration via Software-Hardware Co-Design for Transformer," *IEEE HPCA*, 2022.
[10] Y. Ding, *et al.*, "HAIMA: A Hybrid SRAM and DRAM Accelerator-in-Memory Architecture for Transformer." *ACM/IEEE Design Automation Conference (DAC)*, pp. 1-6, Jul. 2023.
[11] M. Kang, H. Shin, and L. Kim,"A framework for accelerating transformer-based language model on ReRAM-based architecture," *IEEE TCAD*, pp.3026 -3039, Oct. 2021.
[12] M. O'Connor, *et al.*, "Fine-grained dram: Energy-efficient dram for extreme bandwidth systems," *IEEE/ACM MICRO*, pp. 41–54, 2017.
[13] S. Li, "Scope: A stochastic computing engine for dram-based in-situ accelerator," *IEEE/ACM MICRO*, 2018.
[14] I. Thakkar, S.Vatsavai, and V. Karempudi, "High-Speed and Energy-Efficient Non-Binary Computing with Polymorphic Electro-Optic Circuits and Architectures." *GLSVLSI*, pp. 545-550, 2023.
[15] M. H. Najafi, D. J. Lilja, M. Riedel. "Deterministic methods for stochastic computing using low-discrepancy sequences." *IEEE/ACM International Conference on Computer-Aided Design (ICCAD)*, pp. 1-8, Nov. 2018.
[16] D. Wu, et al. "Ugemm: Unary computing architecture for gemm applications." *ACM/IEEE 47th Annual International Symposium on Computer Architecture (ISCA)*. May 2020.
[17] K. Kim, J. Lee, and K. Choi, "Approximate de-randomizer for stochastic circuits," *IEEE ISOCC*, pp. 123-124, Nov. 2015.
[18] S. Shivanandamurthy, et al, "AGNI: In-Situ, Iso-Latency Stochastic-to-Binary Number Conversion for In-DRAM Deep Learning," *ISQED*, 2023.
[19] S. Mysore, I. Thakkar, and S. Salehi, "Atria: A bit-parallel stochastic arithmetic based accelerator for in-dram cnn processing." *IEEE ISVLSI*, 2021.
[20] V. Seshadri, *et al.,* "Ambit: In-memory accelerator for bulk bitwise operations using commodity DRAM technology." *Proceedings of the IEEE/ACM MICRO*, 2017.
[21] J. Ahn, Y. Sungjoo, M. Onur, and C. Kiyoung, "PIM-enabled instructions: A low-overhead, locality-aware processing-in-memory architecture." *ACM SIGARCH Computer Architecture News* 43, pp. 336-348, 2015.
[22] K. Soroosh, Y. Zha, J. Zhang, and J. Li., "Challenges and Opportunities: From Near-memory Computing to In-memory Computing," *ACM ISPD*, pp. 43–46, 2017.
[23] P. Naebeom, R. Sungju, K. Jaeha, and K. Jae-Joon Kim, "High-throughput Near-Memory Processing on CNNs with 3D HBM-like Memory," *ACM TODAES*, Nov. 2021.
[24] M. Lenjani, "Fulcrum: A simplified control and access mechanism toward flexible and practical in-situ accelerators," *IEEE HPCA*, pp. 556-569, Feb. 2020.
[25] F. Gao, G. Tziantzioulis, and D. Wentzlaff, "ComputeDRAM: In-Memory Compute Using Off-the-Shelf DRAMs," *Proceedings of the IEEE/ACM MICRO*, 2019.
[26] Y. Long, T. Na, and S. Mukhopadhyay, "ReRAM-based processing-in-memory architecture for recurrent neural network acceleration," *IEEE VLSI),* pp. 2781-2794, 2018.
[27] X. Qiao, et al., 2018, "AtomLayer: A universal ReRAM-based CNN accelerator with atomic layer computation," *DAC*, pp. 1-6, June 2018.
[28] Y. Chen, "ReRAM: History, Status, and Future*," IEEE Transactions on Electron Devices*, vol. 67, no. 4, pp. 1420-1433, 2020.
[29] V. Seshadri, *et al.*, "RowClone: Fast and energy-efficient in-DRAM bulk data copy and initialization." *IEEE/ACM MICRO*, pp.185-197, 2013.
[30] X. Xin, Y. Zhang, and J. Yang. "Roc: Dram-based processing with reduced operation cycles." *DAC,* pp. 1-6. 2019
[31] S. Vatsavai, I. Thakkar, "A Bit-Parallel Deterministic Stochastic Multiplier," in IEEE ISQED, 2023
[32] C. Yu, *et al.,* "A Logic-Compatible eDRAM Compute-In-Memory With Embedded ADCs for Processing Neural Networks," *IEEE TCAS*, pp. 667-679, Feb. 2021.
[33] S. Afifi, F. Sunny, M. Nikdast, S. Pasricha, "TRON: transformer neural network acceleration with non-coherent silicon photonics." *Proceedings of the Great Lakes Symposium on VLSI 2023* (pp. 15-21), Jun 2023.
[34] M. F. Ali, A. Jaiswal, and K. Roy, "In-memory low-cost bit-serial addition using commodity dram technology," *IEEE Transactions on Circuits and Systems I: Regular Papers*, vol. 67, no. 1, pp. 155–165, 2019.
[35] C. Eckert, *et al.*, "Neural cache: Bit serial in-cache acceleration of deep neural networks," *ACM/IEEE ISCA*, pp. 383–396, 2018.
[36] M. Imani, S. Gupta, Y. Kim, and T. Rosing, "Floatpim: In-memory acceleration of deep neural network training with high precision," *ACM/IEEE ISCA, IEEE*, pp. 802–815, 2019.
[37] HP Labs : CACTI. [Online]: https://github.com/HewlettPackard/cacti.
[38] Y. Li *et al*., "Capacitor-based Cross-point Array for Analog Neural Network with Record Symmetry and Linearity," *ISVLSI*, pp. 25-26, 2018.
[39] N. Kitaev, L. Kaiser, and A. Levskaya, "Reformer: The efficient transformer," *ICLR*, 2020.
[40] L. Siyuan, et al., "Hardware accelerator for multi-head attention and position-wise feed-forward in the transformer," *IEEE SOCC*, 2020